\documentclass[epj, nopacs]{svjour}

\usepackage{amsmath}
\usepackage{latexsym}
\usepackage{graphicx}
\usepackage{hyphenat}
\usepackage{caption}
\usepackage{subcaption}
\usepackage{floatrow}
\usepackage{placeins}
\usepackage{cuted}
\usepackage{makebox}
\usepackage[title]{appendix}
\usepackage{amsfonts}
\usepackage{amssymb}
\usepackage{bm}
\usepackage{microtype}

\begin{document}

\title{Studies on the readability and on the detection rate in a
Mach-Zehnder interferometer-based implementation for high-rate, long-distance QKD protocols}

\author{Christos Papapanos\inst{1}, Dimitris Zavitsanos\inst{1}, Adam Raptakis\inst{1}, Giannis Giannoulis\inst{1}, Christos Kouloumentas\inst{1} \and Hercules Avramopoulos\inst{1}
}                     

\institute{\inst{1}Photonics Communication Research Laboratory,
	School of Electrical and Computer Engineering, National Technical University of Athens, Iroon Polytechniou Street 15780 Greece}

\date{}
\journalname{}
%

\abstract{
We study the way that chromatic dispersion affects the visibility and the synchronization on Quantum Key Distribution (QKD) protocols in a widely-used setup based on the use of two fiber-based Mach-Zehnder (MZ) interferometers at transmitter/receiver stations. We identify the necessary conditions for the path length difference between the two arms of the interferometers for achieving the desired visibility given the transmission distance- where the form of the detector's window can be considered. We also associate the above limitations with the maximum detection rate that can be recorded in our setup, including the quantum non-linearity phenomenon, and to the maximum time window of the detector's gate. Exploiting our results we provide two methods, depending on the clock rate of the setup, to perform chromatic dispersion compensation techniques to the signal for keeping the correct order of the transmitted symbols. At the end, we apply our theoretical outcomes in a more realistic QKD deployment, considering the case of phase-encoding BB84 QKD protocol, which is widely used. Our proposed methods, depending on the transmission distance and on the photon emission rate at transmitter station, can be easily generalized to every fiber-optic QKD protocol, for which the discrimination of each symbol is crucial.
}
\maketitle
\section{Introduction}
\label{intro}
New technological achievements have provided the Quantum Key Distribution (QKD) systems to constantly expand to unprecedented transmission distances \cite{long1,long2,long3,Geneva}; exploiting practical and simple synchronisation techniques at kHz-scale rates \cite{long1}, advanced detection units together with novel finite-key security analysis \cite{long3} and superconducting nanowires combined with ultra-low loss fibers and cold filters to suppress the background noise \cite{Geneva}, successful QKD protocol implementation can be realized through hundreds of kilometers of fiber optics. Moving towards noiseless setup implementations, the authors claim BB84 protocols with fiber distances of ~600 km \cite{Geneva}. Apart from the BB84-based QKD implementations, the new era of emerged protocols such as twin-QKD \cite{Lucamarini}, promise to overcome the rate-distance limitations \cite{Pirandola} by greatly extending the range of secure quantum communications. In this context, other physical limitations that were previously ignored should now be revisited in order to provide a practical implementation framework for this new ultra-long fiber transmission QKD ecosystem.

Single photon sources play a major role in quantum information technologies and their standardization has been widely studied \cite{standard1,standard2}. However in QKD protocols, not only the kind of the components that constitute the QKD setup are responsible for the kind and the efficiency of the protocol created, but the values of the parameters involved too. For example, the phase encoding BB84 and its decoy-state modification QKD protocols have the same configuration; the synchronization and the value of parameters involved are the only difference between them. Hence, the study of the phenomena that take place in optical fibers for the specific setup is crucial for maximizing the efficiency of each protocol.

\begin{figure}[h!]
	\centering
	\includegraphics[width=\linewidth]{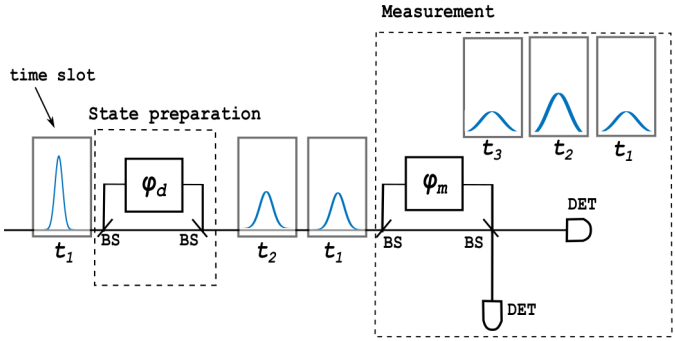}
	\caption{Orthogonal (ideal) time slot implementation of QKD protocols. The three pulses at the output are shown widened compared to the incoming pulse, due to the effect of chromatic dispersion on photon qubits. BS, beamsplitter (50/50); DET, detector.}
	\label{fig: time_bin_encode}
\end{figure}

Many QKD protocols- including the aforementioned ones- use time bin encoding for the creation and the reception of the qubit. This method is advantageous for long-distance quantum communication \cite{Diamanti}. The setup used for time-slot implementation of prepare and measure protocols is depicted in Fig. \ref{fig: time_bin_encode}; the incoming pulse at the input is divided into three pulses at the output. The components with more details are presented in Fig. \ref{fig:two_mach}. We will later prove that the exterior pulses do not contain any information for the transmitted states and so they should be distinguished from the middle pulse to avoid the temporal overlapping. As a result, the correct synchronization is important for reading the middle pulse (indicated by time slot \(t_2\)). It is evident that the temporal width of the pulse plays a crucial role to prevent the temporal overlapping as well as to obtain the proper synchronisation. 
For these two reasons, chromatic dispersion of the fiber medium has a major impact on the protocol's efficiency, especially as the communication distance is significantly increased.

The added value of our contribution is the definition and the theoretical establishment of a complete bottom-up approach to the natural limitations created by chromatic dispersion and the countermeasures that should be taken to confront its effect- apart from the ordinary use of Dispersion Compensating Fibers (DCF)- for maintaining the correct sequence of the transmitted pulses (keeping a low cost) and maximizing the readability and the detection rate of the setup. Beginning from a theoretical establishment and examination of the model developed in \cite{CD}, we find new limits and restrictions; lastly we present an engineering approach over the practical use of our theoretical conditions that our results can provide.

In our paper, it is proved that the proper choice of the phase shift values of the two Mach-Zehnder (MZ) interferometers is sufficient and necessary to 
provide the output with the required readability, despite the effect of chromatic dispersion, for performing decoding and synchronization. 

More specifically, the condition (lower bound) that needs to be satisfied between the phase shifters of the two interferometers for a given fiber distance between Alice (transmitter) and Bob (receiver) stations in order to attain a specific visibility value is extracted. Furthermore, the form of the detector's reading window is taken into consideration, leading to a more realistic model.

The theoretical maximum detection rate (raw key rate) that can be generated is related to these setup parameters and, thus we are also able to identify an upper bound. As it was originally expected, a trade-off between the visibility of the setup and the theoretical maximum detection rate, and thereby with the distilled secure key is revealed. This upper bound directly indicates the form of the synchronization- when and for how long Bob needs to read each pulse- that needs to exist for properly reading the signal.

At longer fiber distances, the widening of each pulse caused by chromatic dispersion effect may lead some symbols to be detected in different order than the way they were sent (i.e. transmitted symbols in order A, B, C might be received in order B, A, C). As a result, a part (Sec. \ref{sec: 7}) of this research is devoted to the definition of a mitigation strategy to overcome the impact of fiber dispersion in our setup. More specifically, we use the estimated maximum detection rate to obtain the minimum required compensation length to preserve the correct arrival order. Two methods are presented depending on the systems pulse repetition rate at Alice station, which can be easily expanded out of this specific two MZ interferometers setup.

During this process, an alternative approach for establishing the values of the phase shifters, necessary for creating the two bases of phase encoding BB84 QKD protocol, is presented in Sec. \ref{appendix A}; where we have used a mathematically original quantum approach. An example is finally presented where our previous theoretical results are used for a better realization of their feasibility.


\begin{strip}
	\floatbox[{\capbeside\thisfloatsetup{capbesideposition={right,top},capbesidewidth=4.9cm}}]{figure}[\FBwidth]
	{\caption{Two MZ interferometers in series: 1 input a, 3 outputs h,o,p, 4 beam splitters (BS), 5 optical fibers (i=c,d,g,m,n), fiber lengths \(l_i\), phase factors \(P_i=exp{\{-ik\varDelta_i\}}\), \(\varDelta_i\) are phase shifters, transmission coefficients \(T_i=exp{\{-2l_i a_i\}}\) (\(a_i\) are absorption factors); position distributions (or time pulses) are drawn. \protect\cite{CD}}\label{fig:two_mach}}
	{\includegraphics[width=12.7cm]{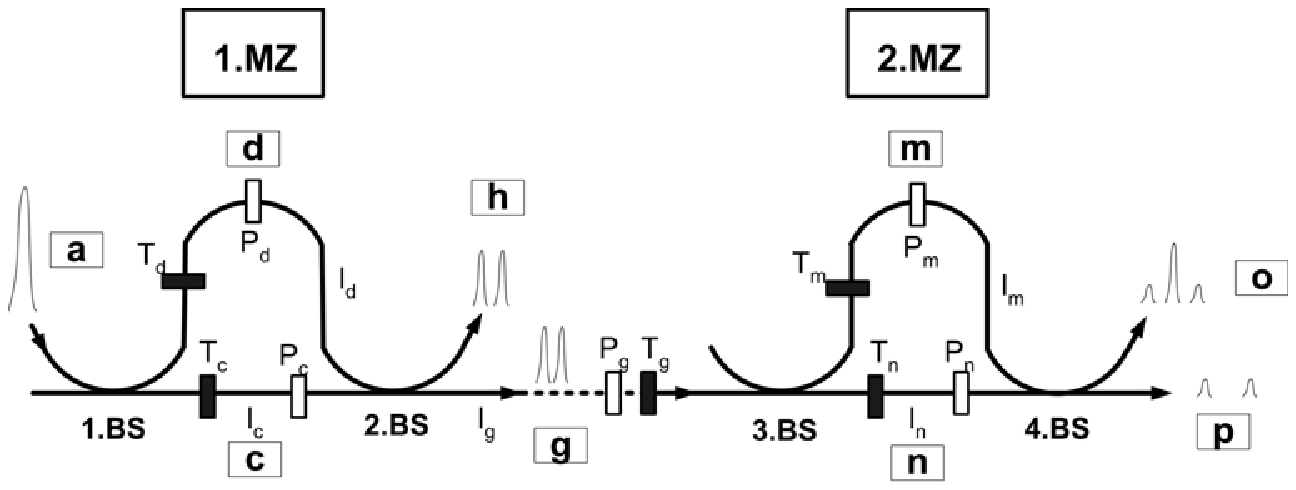}}
\end{strip}

\section{Definition of the model for QKD protocols}
\label{sec:1}
Chromatic dispersion causes the widening of the transmitted pulses with a temporal intensity profile of Gaussian shape. In \cite{CD}, the results of the transmission of a Gaussian pulse through two MZ interferometers, connected in series as indicated in Fig. \ref{fig:two_mach}, have been studied. Gaussian pulses have been chosen as being mathematically convenient and useful approximation to true pulse shapes. When the pre-compensation technique is adopted the pulse inserted in the setup is not Gaussian but our paper’s results do not alter (discussed in Sec. \ref{sec: 5} and appendix \ref{appendix B}). We make use of the mathematical formalism presented in \cite{CD}, which is true for the general case regarding the possible parameters that characterize the interferometers, as a tool to provide applicable, ready to use formulas. As a result, we have chosen to use the same symbols, see Fig. \ref{fig:two_mach}, for our formalism.

In our paper, we consider the case that is true for most QKD protocols, which uses two MZ interferometers (ex. phase encoding BB84 QKD, decoy-state phase encoding QKD). In these protocols, the setup uses two identical asymmetric MZ interferometers with the only difference being the phase factors in their long paths (\(\varDelta_d\neq \varDelta_m\)). The difference between these values is, almost always, of the order of wavelength; more specifically (\(0\), \(\lambda_0/4\), \(\lambda_0/2\), \(3\lambda_0/4\)) which correspond to phase shift values (\(0\), \(\pi/2\), \(\pi\), \(3\pi/2\)). However, our model is true for the general case where \(\varDelta_d- \varDelta_m\) can take any value.
Furthermore, in most QKD protocols Gaussian pulses are transmitted through the same fiber link between Alice and Bob and thus the final output depends on phase difference. Hence, the effect of $\varDelta_g$ is immaterial and hereinafter we consider \(\varDelta_g=0\).

From \cite{CD} we can extract the following result for the position spectra at the exits o, p:

\begin{equation} \label{position spectra}
\centering
	\begin{split}
	|\psi_{o,p}(x)|^2 =&\dfrac{T_g}{32\pi\sqrt{2\pi}(\delta k)} \{|J_{dm}(x)|^2+|J_{cm}(x)|^2\\
	&+|J_{dc}(x)|^2+|J_{cc}(x)|^2+2II_{o,p}(x) \}
	\end{split}
\end{equation}

where

\begin{equation} \label{II}
	\centering
	\begin{split}
	II_{o,p}(x)= &- \Re[J_{dm}(x) J^*_{cm}(x)] \mp \Re[J_{dm}(x) J^*_{dc}(x)]\\ 
	&\pm \Re[J_{cm}(x) J^*_{dc}(x)] \pm \Re[J_{dm}(x) J^*_{cc}(x)]\\
	& \mp \Re[J_{cm}(x) J^*_{cc}(x)] - \Re[J_{dc}(x) J^*_{cc}(x)]
	\end{split}
\end{equation}
where in both Eqs. \ref{position spectra} and \ref{II} we considered the restrictions of the previously described setup where \(c\equiv n\) (c, n represent the short legs of each MZ as shown in Fig. \ref{fig:two_mach}).

We have chosen the squared absolute value of the position wave function, \(\psi_{o,p}(x)\), as it represents the position spectrum and has the desired meaning of the probability that we will later use.

In Eqs. \ref{position spectra} and \ref{II} the single quantities are defined \cite{CD}:

\begin{equation} \label{more equations}
\centering
	\begin{split}
	|J_{ij}(x)|^2 =& 4\pi (\delta k)^2 \sqrt{\dfrac{1}{\gamma_{ij}}} T_i T_j \\
	 &\times exp\{-2(\delta k)^2[x'_{ij}-2\delta_{ij} k_0]^2/ \gamma_{ij} \}\:, \\		
	\Re[J_{ij}(x) J^*_{kl}(x)] =& c'_{ij} c'_{kl} cos[z_{ij}(x)-z_{kl}(x)]\:,\\
	c'_{ij}(x)=&2(\delta k)\sqrt{\pi}\sqrt{T_iT_j}(\gamma_{ij})^{-1/4} \\
	&\times exp\{-(\delta k)^2[x'_{ij}-2\delta_{ij} k_0]^2/ \gamma_{ij} \}\:,\\
	z_{ij}(x) =&\frac{1}{2}arctan[4\delta_{ij}(\delta k)^2]\\
	&+\dfrac{1}{\gamma_{ij}}[k^2_0\delta_{ij}-k_0x'_{ij}-4(\delta k)^4\delta_{ij}{x'_{ij}}^2]
	\end{split}
\end{equation}

In paper \cite{CD}, it is defined:

\begin{equation} \label{delta}
	\centering
	\begin{split}
	\delta_{ij}=B_g+B_i+B_j
	\end{split}
\end{equation}
where \(B_i=\kappa l_i\) with \(\kappa=-D_{\lambda_0}\lambda_0^2c_0/(4\pi)\). $D_{\lambda_0}$ is the dispersion coefficient which equals to $17\:\frac{ps}{km\cdot nm}$ for $\lambda_0=1550\:nm$. We can see that because of our restrictions, every possible combination of i,j gives \(\delta_{ij}\) the same value. Therefore, from now on we will call every pair \(\delta_{ij} \equiv \delta_1\) (not to be confused with \(\delta k\)).

We also have:

\begin{equation} \label{gamma}
\centering
	\begin{split}
	\gamma_{ij}&=1+16(\delta k)^4{\delta_{ij}}^2 \\
	&=1+16(\delta k)^4{\delta_1}^2
	\end{split}
\end{equation}
where \((\delta k)^2\) is the mean square deviation of wave numbers of the input Gaussian pulse/function and it holds \(\delta k=2\pi (\delta \lambda) /{\lambda_0}^2\). In the same way, we find that the parameter \(\gamma_{ij}\) has the same value for every combination of i, j and, thereby we will call it \(\gamma\).

It is defined:

\begin{equation} \label{eqn: x'_ij}
	\centering
	\begin{split}
	x'_{ij}&=x-A_g-A_i-A_j
	\end{split}
\end{equation}
where $x$ is the total distance that the input pulse has propagated until the outputs of the MZ interferometer and \(A_i=\varDelta_i+N_0l_i\), where \(N_0=N(\lambda_0)\) is the group index for the mean wavelength \(\lambda_0\). As mentioned in \cite{CD}, \(x\) is the (hypothetical) distance as if the photons have the vacuum speed of light. Furthermore, using the same logic the parameters \(T_i,\: l_i\) can be expressed uniquely as \(T,\: l\) respectively.

\section{Interpretation of the model}
\label{sec:2}
In order to interpret the previously established model, we define the parameter:

\begin{equation} \label{mu}
\centering
	\begin{split}
	\mu_{ij}&=N_0l_g+\varDelta_i+\varDelta_j+2N_0l+2\delta_1 k_0
	\end{split}
\end{equation}

One important observation worth mentioning at this point is that although \(\mu_{ij}\) differs for every pair ij, the difference of the values is very small due to the fact that 
 \(l_g\gg \varDelta_i\) for every i and for long transmission distances.

Finally, we can express more clearly the Eqs. \ref{more equations} in the following form:

\begin{equation} \label{more equations final}
	\centering
	\begin{split}
	|J_{ij}(x)|^2 =& 4\pi (\delta k)^2 T^2\sqrt{\dfrac{1}{\gamma}} \\
	 &\times exp\{-2(\delta k)^2[x-\mu_{ij}]^2/ \gamma \}\:,
	\\		
	\Re[J_{ij}(x) J^*_{kl}(x)] =& c'_{ij} c'_{kl} cos[z_{ij}(x)-z_{kl}(x)]\:,
	\\
	c'_{ij}(x)=&2(\delta k){\gamma}^{-1/4}T\sqrt{\pi} \\
	&\times exp\{-(\delta k)^2[x-\mu_{ij}]^2/ \gamma \}\:,
	\\
	z_{ij}(x) =&\frac{1}{2}arctan[4\delta_1(\delta k)^2]\\
	&+\dfrac{1}{\gamma}[k^2_0\delta_1-k_0x'_{ij}-4(\delta k)^4\delta_1 {x'_{ij}}^2]
	\end{split}
\end{equation}

As can be now seen, \(|J_{ij}(x)|^2,\: \Re[J_{ij}(x) J^*_{kl}(x)]\) are Gaussian distributions with a mean value given by:

\begin{equation} \label{eqn: mean value}
\centering
		\begin{aligned}
			\mu_1=& \mu_{ij}\\
			\mu_2=& \dfrac{\mu_{ij} + \mu_{kl}}{2}
		\end{aligned}	
\end{equation}

respectively. The standard deviation is given by:

\begin{equation} \label{eqn: sigma}
\centering
	\begin{aligned}
	\sigma=\frac{\sqrt{\gamma}}{2(\delta k)}
	\end{aligned}	
\end{equation}

for both distributions. Obviously, the cosine term on Eq. \ref{more equations final} changes the amplitudes of each Gaussian distribution.

For computing the mean value and the standard deviation of the second term (\(\Re[J_{ij}(x) J^*_{kl}(x)]\)), the formulas of the product of two Gaussian distributions with arbitrary means \(\mu_{f}, \mu_{g}\) and standard deviations \(\sigma_{f}, \sigma_{g}\) have been used.

Now, we are in position to express the full width half maximum (FWHM) value of each pulse. The FWHM is the same for every pulse since it depends on the variance and not on the mean value. So:

\begin{equation} \label{eqn: FWHM}
	\centering
	\begin{aligned}
	FWHM&=\sqrt{8\ln2} \cdot \sigma 
	\end{aligned}	
\end{equation}

In conclusion, the signal at the MZ interferometer outputs set prior to the photon detectors, is a sum of Gaussian distributions with the same standard deviation and slightly different mean values.

\section{Results and Discussion}
\label{sec: 3}
The range of the derived FWHM is not a sufficient metric for QKD protocols, as we will later see; neither is the wider one, where the energy of the pulse falls to the \(1/e\) of its maximum value. The half width of the latter value is often used in classical description of Gaussian pulses and is symbolized \(T_0\). As a result, a wider metric is needed.

For determining the proper width of this metric, we need to define the new parameter \(X_\varrho\) which is the half width of the pulse where its energy falls to \(\frac{1}{e^{\varrho^{2}}}\).

\begin{equation} \label{eqn: Xk}
\centering
	\begin{aligned}
	X_\varrho&=\varrho\dfrac{FWHM}{2\sqrt{\ln2}} \\
	&=\varrho \cdot \sqrt{2}\cdot\sigma
	\end{aligned}	
\end{equation}

As expected for \(\varrho = \sqrt{ln2}\) we have \(X_{\sqrt{ln2}}=FWHM/2\) and for \(\varrho = 1\) we have \(X_{1}=T_0\).

\subsection{Lower bound - High visibility}
\label{sec: 4}
Achieving high visibility demands the complete knowledge of the signal at the MZ interferometer outputs. Since every pulse at the output of the fiber setup exhibits the same spreading, as a next step we need to find the longest distance between any two Gaussian distributions at the output of the setup produced by one incoming pulse.

From Eqs. \ref{more equations final} and \ref{eqn: mean value}, we derive that the maximum possible distance between two Gaussian distributions of Eq. \ref{position spectra} depends only on the values of the phase shifters, \(\varDelta_d\), \(\varDelta_m\) and \(\varDelta_c\). The phase shifting can be applied by extending the corresponding lengths of the fibers as mentioned in \cite{CD} while the phase modulation can be implemented by employing high-speed phase shifters which are capable to introduce phase shifting of the order of a few wavelengths (see Sec. \ref{appendix A}). The longest distance between the Gaussian distributions is between the part of the pulse that propagates through the short-short legs of the two MZ interferometers (\(|J_{cc}(x)|^2\) term) and the part of the pulse that propagates through the long-long legs of the two MZ interferometers (\(|J_{dm}(x)|^2\) term). It is evident that this distance is the same for both \(|\psi_{o}(x)|^2\) and \(|\psi_{p}(x)|^2\) because the aforementioned terms exhibit the same dependence on the exterior Gaussian distributions.

\begin{figure}
	\begin{subfigure}[t]{\columnwidth}\centering
		\includegraphics[width=8.8cm]{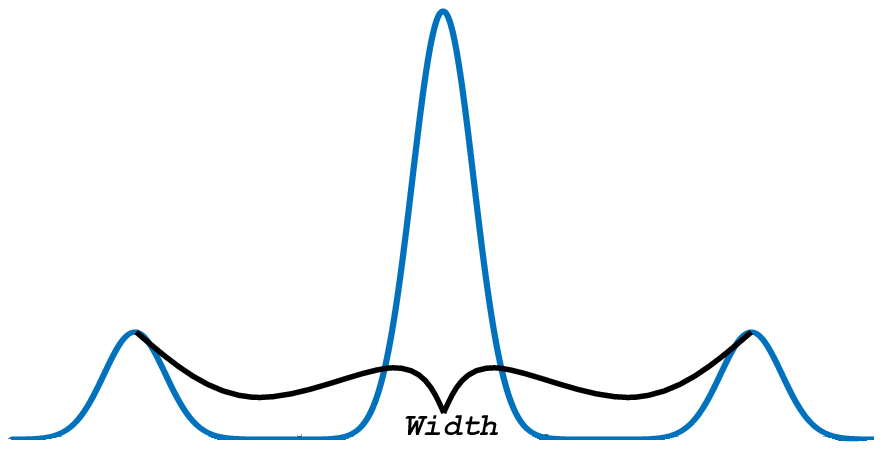}
		\subcaption{Width of one symbol (considering only the mean value).}
		\label{fig: width}
	\end{subfigure}
	\begin{subfigure}[t]{\columnwidth}\centering
		\includegraphics[width=8.8cm]{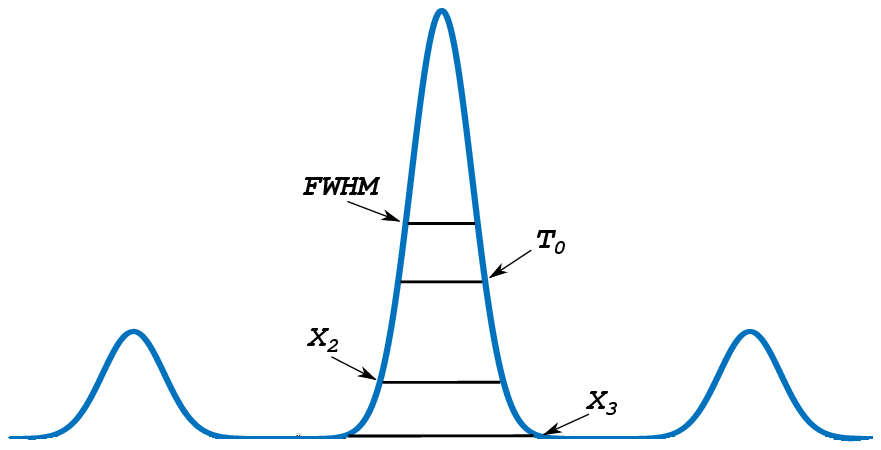}
		\subcaption{The whole range of parameters \(2\cdot X_\varrho\) on a symbol.}
		\label{fig: Xdiafora}
	\end{subfigure}	
	\caption{Overview of the shape of a symbol at the MZ interferometer outputs.}
	\label{fig: symbol}
\end{figure}

Subtracting the distances that the mean value of the aforementioned two pulses have propagated along the entire fiber link (meaning \(\mu_{cc}\), \(\mu_{dm}\)), we find the maximum width of the exterior Gaussian distributions. If we remove the standard deviation (\(\sigma\)) of each Gaussian distribution, the final width of the symbol because of chromatic dispersion effect is shown in Fig. \ref{fig: width}.
Recalling that \(\mu_{dm}>\mu_{cc}\), this width is equal to:

\begin{equation} \label{eqn: difference}
	\centering
	\begin{aligned}
	(x-\mu_{cc})-(x-\mu_{dm}) &= \varDelta_d+\varDelta_m-2\varDelta_c \\
	&= \varDelta_d+\varDelta_m
	\end{aligned}	
\end{equation}
where we replaced \(\varDelta_c =0\), a consideration that most of the similar setups meet or consider it as a benchmark. The terms $x-\mu_{cc}$ and $x-\mu_{dm}$ express the distance that the two exterior pulses have propagated in their mean value frame.

Figure \ref{fig: Xdiafora} shows an overview of the different values of \(X_\varrho\). It is worth mentioning that these values are identical for each pulse, including the middle one and the two exterior ones. Since the exterior pulses are present in every case, they do not provide us any information about the basis of the signal; as a result we need to separate them from the middle pulse. We can see that as the width (Eq. \ref{eqn: difference}) decreases, a critical point will be approached. Beyond this point, the 3 pulses will overlap and, thus we will not be able to decode the information. We are then able to extract the necessary condition to achieve this separation, given the visibility that we want to succeed:

\begin{equation} \label{eqn: accuracy}
	\centering
	\varDelta_d+\varDelta_m\geq4X_\varrho=2\varrho\dfrac{FWHM}{\sqrt{\ln2}}=\varrho \cdot 4\sqrt{2}\sigma	
\end{equation}
where $\varrho$ expresses the visibility in the way that will be mentioned in the next paragraph.

\FloatBarrier

As stated above, each pulse is a Gaussian distribution and it is also the position spectrum of the particle that the symbol contains. Hence, we are able to interpret it as the probability of detecting the particle in each position, so by transforming the Gaussian distributions of the pulses to standard normal distributions we can estimate the detection probability in each pulse in relevance with the parameter \(\varrho\). A similar approach could be used for other than Gaussian pulses as the input pulse. 

Table \ref{table:k and probability} summarizes the main results that can be provided from Eq. \ref{eqn: accuracy}. We have considered only the equality of this condition and furthermore we can conclude that there is no point of selecting values of $\varrho$ higher than 3 for the below reasons:

\begin{enumerate}
	\item We accomplish the maximum probability for correct distinction which equals to 1.
	
	\item When the phase shifting is created having as a corollary the enlargement of the lengths of the corresponding MZ legs, then as $\varrho$ increases, the length of the overall transmission distance also increases. Leading to higher losses (even if the transmission distance increase is of the order of meters) and, thus to lower detection rate.
	
	\item A more important factor is the maximum detection rate which is restricted from the intersymbol interferences. We will analyze this factor in detail in the next subsection.
\end{enumerate}

\begin{table}[h] \centering
	\begin{tabular}{|c|c|c|}
		\hline
		\(\varrho\) & \(\varDelta_d+\varDelta_m\) & Visibility  \\ \hline
		1 & \(2.402\cdot FWHM\)           & 84.26 \%   (1.414\(\cdot\sigma\))     \\ \hline
		2 & \(4.804\cdot FWHM\)           & 99.54 \%   (2.828\(\cdot\sigma\))     \\ \hline
		3 & \(7.206\cdot FWHM\)           & 100 \%   (4.243\(\cdot\sigma\))       \\ \hline
	\end{tabular}
	\caption{Visibility of the signal, depending on the path difference of the two Mach-Zehnder interferometers.}
	\label{table:k and probability}
\end{table}

The probabilities which are presented in table \ref{table:k and probability}, in the case of the ideal configuration (except, of course, for the chromatic dispersion), are the values of the interference Visibility of the setup. Visibility is an important parameter since it is directly connected with the Quantum Bit Error Rate (QBER) performance which in turn assesses- the secure key rate that can be distilled \cite{Branciard_2005,Eraerds_2010}.

In case that there are also other imperfections- for example not perfect alignment- on the setup (referring only on the line between Alice and Bob), the total Visibility of the setup can be estimated as the product of the individual probabilities for the case of independent variables (independent imperfections). In practical QKD installations, the Visibility ranges from \(93\%\) to \(99.999\%\). \cite{Lucamarini}

At this point, we need to mention that the above consideration of the equality (Eq. \ref{eqn: accuracy} and Table \ref{table:k and probability}) considers the existence of an ideal gated detector; meaning the rising and falling times are zero and, thus the shape of the time slot that the detector reads is orthogonal as shown in Fig. \ref{fig: time_bin_encode}. Gated detectors are more effective than free-running detectors when short optical pulse trains have to be detected, as in QKD applications \cite{detectors}. In case where these characteristics of the detectors should be taken into consideration, the shape of the time slot is changed to trapezoid and the right term of the condition of Eq. \ref{eqn: accuracy} is increased by the rising and falling times, which have been expressed in units of length. More specifically, considering that the detectors could read with precision only at the plateau of the time slot, we obtain: 

\begin{equation} \label{eqn: accuracy2}
\centering
	\begin{aligned}
	\varDelta_d+\varDelta_m\geq & 4X_\varrho+c_0\cdot (t_{rising}+t_{falling}) \\
	= & 2\varrho\dfrac{FWHM}{\sqrt{\ln2}}+c_0\cdot (t_{rising}+t_{falling}) \\
	= & \varrho \cdot 4\sqrt{2}\sigma+c_0\cdot (t_{rising}+t_{falling})
	\end{aligned}		
\end{equation}
where $c_0$ is the speed of light in the vacuum. An effective index of the transmitted mode is not needed because, as we have already mentioned, the initial formulas define \(x\) as if photons had the vacuum speed of light. In general, in Eq. \ref{eqn: accuracy} a safety factor could be considered.

\begin{figure}[h!]
	\centering
	\includegraphics[height=8.8cm]{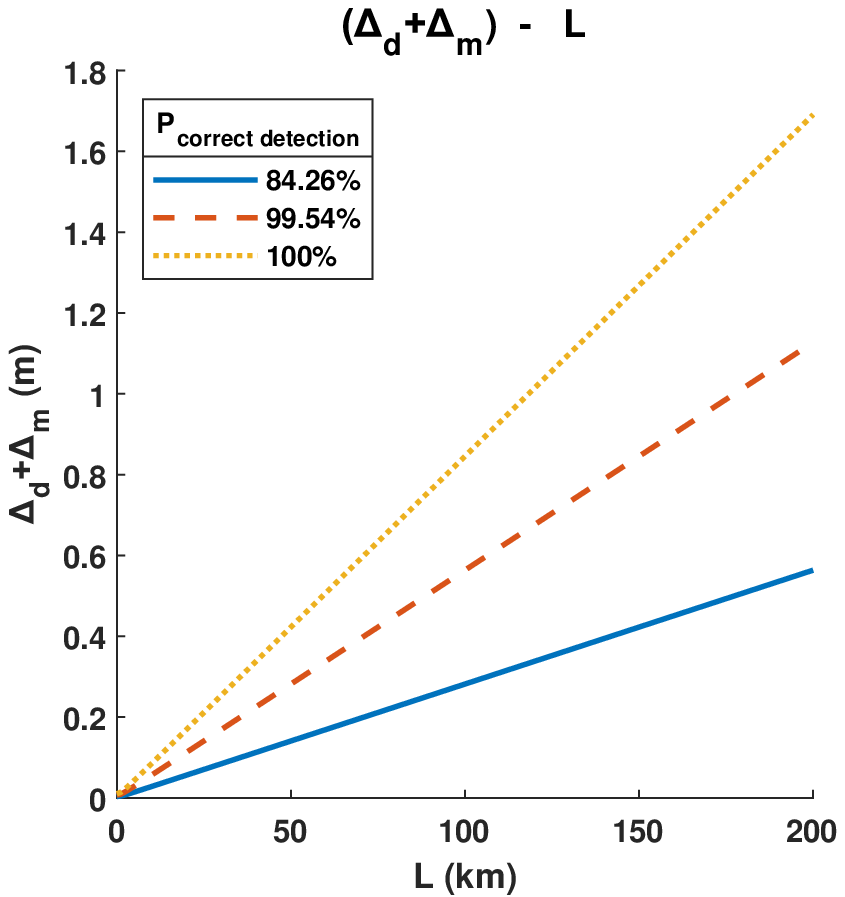}
	\caption{Minimum sum of the values of the phase shifters needed because of chromatic dispersion.}
	\label{fig: Sum of D}
\end{figure}

The results that we have extracted can be applied to QKD implementations relying on the use of MZ interfero-meters-assisted state preparation and measurements stations. Hence, they are universal and in Fig. \ref{fig: Sum of D} are presented in the range that this category of QKD protocols operate. As it was originally expected, they appear a linear dependence on distance and the value of the sum of the phase shifters is increased for higher detection rates.

It is worth applying realistic \(t_{rising}\) and \(t_{falling}\) times to Eq. \ref{eqn: accuracy2} in order to define the dominant term of this condition. In \cite{response time} approximate response times for different superconducting nanowire single-photon detectors (SNSPDs) are given; the response time consists of the rise time and fall time ($t_{response}=t_{rise}+t_{fall}$). SNSPDs can exhibit response time around to \(5\:ns\). Hence, the added term coming from the imperfect features of the detector equals to \(1.5\:m\) and the first sum as indicated in Fig. \ref{fig: Sum of D} with the dotted line is around \(1.8\:m\) for \(200\:km\) transmission distance; for longer distances the gap between those terms is widened.

As a result, advanced technology establishes the second term to be of the same order as the first term leading to higher phase shift values. However, it is evident that when such SNSPDs are not available, the response time will have a more dominant role. 

Finally, practical conditions for high quality readability have been defined and the proper metric for QKD systems that we were looking for is \(X_3\). This metric can be used beyond this two MZ interferometers setup.

As a next step, we will analyse the third important aforementioned factor that prevents the intersymbol interference and maintains the correct arrival order of each symbol (this interpretation is presented in Sec. \ref{sec: 7}).

\subsection{Upper bound - Theoretical maximum detection rate}
\label{sec: 5}
At the output of the setup, owing to the chromatic dispersion each symbol has its widest pulse width and then, after a very short distance (from the MZ's output to the detector), it will be detected. We will symbolize this distance as MZ-D. Each symbol, as we have mentioned, is consisted of three pulses. The pulses which come from the short-short travel, the short-long/long-short travel and the long-long travel of the symbol will be from now on referred as right one, middle one and left one respectively, where we have considered, without loss of generality, that the symbol propagates along the right axis. Considering two consecutive pulses, there is a distance threshold above which these two consecutive symbols will start to interact. The parts of the symbols that will firstly interact are the left pulse of the first symbol and the right pulse of the second symbol.

As already mentioned, the information is hidden in the middle pulse of each symbol and, thus we need to find a condition that will prevent the interaction of the two consecutive symbols being able to affect the middle pulse/window altering the order and/or the value of the symbol read. 

When the two symbols start to temporally overlap, the quantum non-linearity phenomenon could possibly arise \cite{quantum_non_linearity,quantum_non_linearity2}. In this case, single photons at new frequencies can be generated and due to chromatic dispersion they will then propagate along fiber with different velocities. Hence, during the distance MZ-D, if the new frequencies are sufficiently fast to cover the adjacent temporal window set prior to the detectors, a part of energy of the interacting pulses could probabilistically leak into the middle pulse, degrading thereby the readability of our setup. However, as can be derived from \cite{quantum_non_linearity,PhysRevA.34.4929}, this is quite impossible to happen at these energy levels and in the conventional telecom fibers testbeds which are currently used in long-distance deployments. Through the literature, there are limited works demonstrating photon-to-photon non-linear interaction using specialty fiber setups where one can probe this quantum non-linear domain \cite{nonlinear}. 

Having this in mind and targeting to keep our results as general as
possible in support of the future experimental realizations, we have
studied both cases; considering and ignoring the quantum non-linearity effect. To this end, we also derive the modified condition for the case where non-linear photon-to-photon interaction can be obtained between the symbols in our model which leads to lower detection rates.

\begin{figure}[h!]
	\begin{subfigure}{\linewidth}
		\includegraphics[height=4.4cm,width=8.8cm]{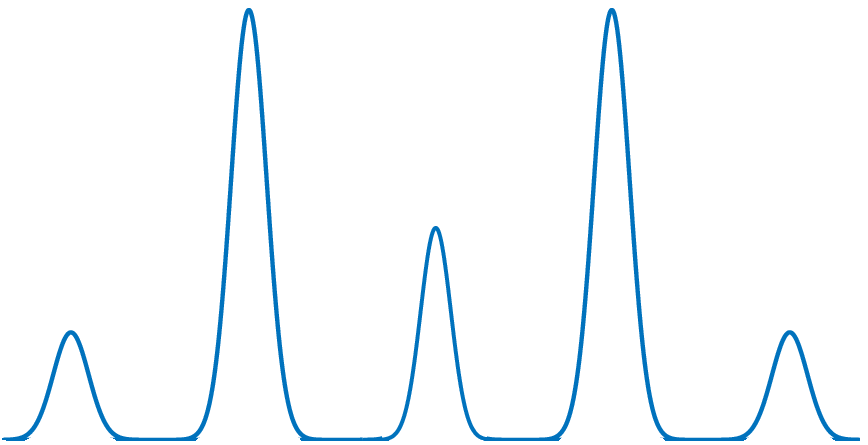}
		\caption{Overview of the minimum distance in the case that the quantum non-linearity phenomenon can be ignored.}
		\label{fig: no4mixing}
	\end{subfigure}
	\begin{subfigure}{\linewidth}
		\includegraphics[height=4cm,width=8.8cm]{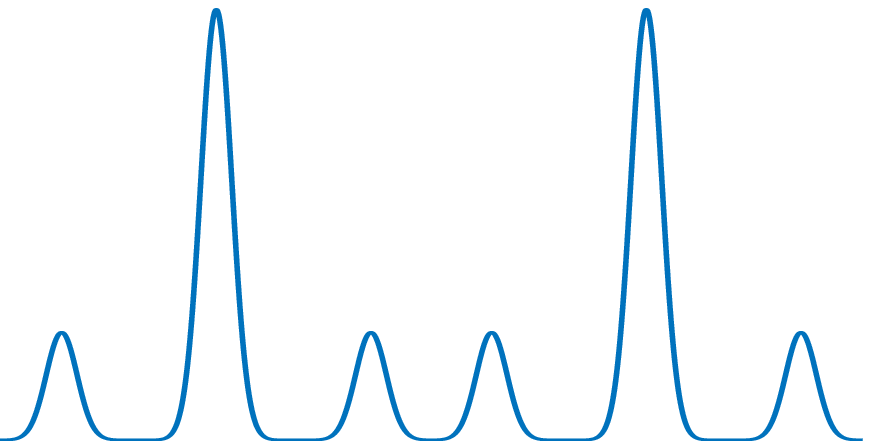}
		\caption{Overview of the minimum distance in the case that the quantum non-linearity phenomenon cannot be ignored.}
		\label{fig: 4mixing}
	\end{subfigure}	
	\caption{Overview of the minimum distance between two consecutive symbols at the MZ outputs in order for the symbols to not temporally overlap. The x-axis indicates the length and the y-axis is pseudo-normalized for the relative height of each pulse to be visible.}
	\label{fig: intersymbol}
\end{figure}

In case where quantum non-linearity phenomenon cannot affect, as previously described, the middle pulses of these two consecutive symbols, we can extract the condition presented in Eq. \ref{eqn: R}. As shown in Fig. \ref{fig: no4mixing}, the distance between two successive symbols needs to be equal to two times the value of the width of the pulse, meaning \(4X_\varrho\), in order to not have intersymbol interference. This straightforwardly leads to:

\begin{equation} \label{eqn: R}
\centering
	\begin{aligned}
	R\leq\dfrac{c}{4X_\varrho}
	\end{aligned}		
\end{equation}
where $R$ is the theoretical maximum detection rate if no countermeasures are taken (see Sec. \ref{sec: 7}), thus an increase of \(\varrho\) leads to a decrease of the maximum possible detection rate. This is the mechanism behind the obtained trade-off between the visibility and the maximum detection rate, as it has been already emphasized in the introduction.
Certainly, the values of the maximum detection rate are higher than those that most protocols have succeed because of the low energy of the pulse and the speed of the electronic systems that the setup uses.

In case where the quantum non-linearity phenomenon could affect the middle pulse, we modified our condition as follows:

\begin{equation} \label{eqn: R2}
\centering
	\begin{aligned}
	R\leq\dfrac{c}{6X_\varrho}
	\end{aligned}		
\end{equation}
so that each symbol has no shared part with any of the consecutive pulses (see Fig. \ref{fig: 4mixing}). As expected from Figs \ref{fig: no4mixing} and \ref{fig: 4mixing}, the theoretical maximum detection rate is smaller when quantum non-linearity has an effect than when it has not. The exact ratio is $1.5$ times smaller as indicated in Eq. \ref{eqn: R2}.

All figures below represent the case that quantum non-linearity has no effect on the middle pulse but using Eq. \ref{eqn: R2}, someone can easily represent the opposite case.

It is worth mentioning that in order to keep the maximum possible detection rate, we should not consider, as in classical channels, the range of the pulse where its power reaches the \(\frac{1}{e}\) of the maximum value but instead the point that its power reaches the \(\frac{1}{e^9}\) of the maximum value. Therefore, the definition of the dispersion length ($L_D$), as defined in \cite{disp_length}, for the quantum setup implementations studied for most QKD protocols, representing the maximum optical fiber length so that the chromatic dispersion will not affect the signal decoding needs to be re-considered and/or re-expressed by future works.

The condition in Eq. \ref{eqn: R} leads to an upper bound for the detection rate that can be obtained, if no countermeasures are taken, and it only depends on the nature of this setup. In \cite{mlejnek2018modeling}, by following a more classical optics approach, it was found that pre-compensation is more preferable for countermeasuring the dispersion-induced readout errors compared to the post-compensation effect and compensation that happens within the fiber link. However, in the aforementioned paper the existence of the exterior pulses has not been taken into consideration, leading to a higher detection rate, but as far as the countermeasure is concerned the results are not affected. Hence, pre-compensation is more preferable.

When compensation techniques are applied, Eqs. \ref{eqn: accuracy} and \ref{eqn: accuracy2} are calculated considering as \(l_g\) to be the active length. In this case, where compensation techniques are applied, Fig. \ref{fig: Sum of D} designates the active length. By the term active length, we describe the part of the length of the optical fiber that has not been compensated; for example, when the transmission length is fully compensated, the active length is zero. In appendix \ref{appendix B}, it is proved that compensation using DCF is possible with all three aforementioned techniques for this specific setup without any change to our results.

At this point, we are able to understand the form of synchronization. Assuming that $\Delta t_{gate}$ is the time window that our detectors are able to read, it needs to fulfil the following condition:

\begin{equation} \label{eqn: synchronization}
\centering
\Delta t_{gate} \leq \dfrac{\varDelta_d+\varDelta_m- 2X_{\varrho}}{c}
\end{equation}
and to be fully synchronised in the middle of the pulse with a reading repetition rate equal to $R$.

\begin{figure}[h]
	\centering
	\includegraphics[height=8.8cm]{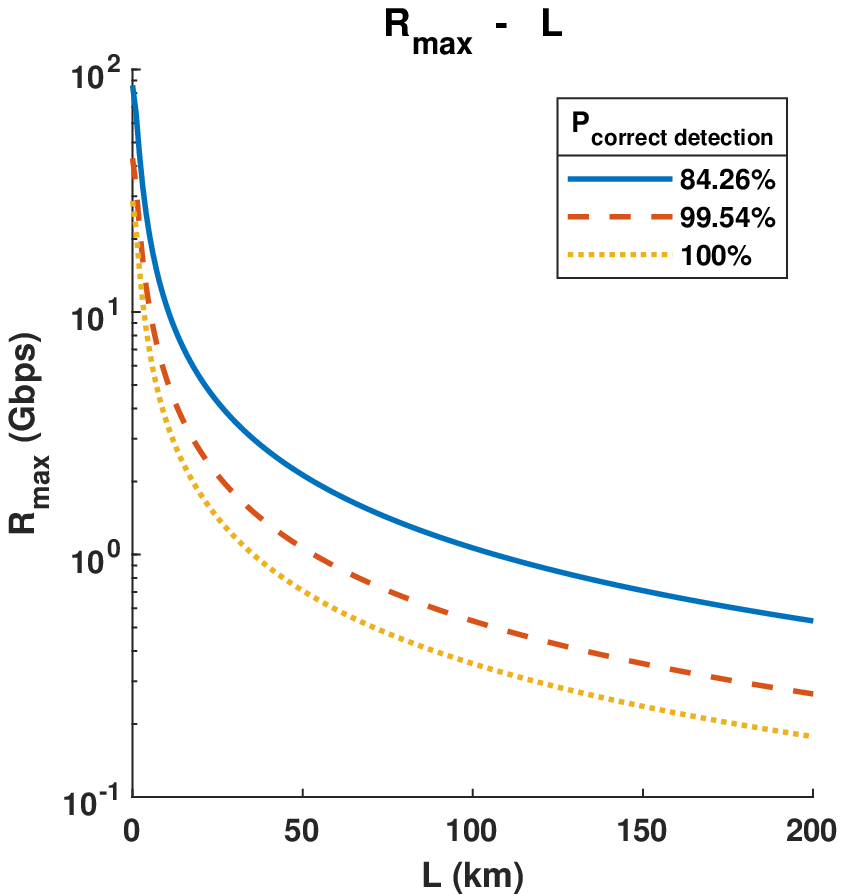}
	\caption{Maximum detection rate creation because of chromatic dispersion when quantum non-linearity phenomenon has no effects. (y-axis is in logarithmic scale)}
	\label{fig: Rmax}
\end{figure}

Below we present some results which indicate the maximum detection rate creation that can be accomplished at Bob station without using any dispersion compensating fiber which increases the distance and, thus the photon losses (Fig. \ref{fig: Rmax}). In all of our simulations, we have considered wavelength equal to \(\lambda_0=1550\:nm\) and divergence from the central wavelength equal to \(\delta\lambda=0.31\:nm\), meaning \(\dfrac{\delta\lambda}{\lambda_0}=2\cdot 10^{-4}\). Our results are true for both deterministic single-photon sources where an ideal emitter can produce one photon at a time and also for probabilistic sources realized through an attenuated laser source. The only characteristics of the source which we have considered are the central wavelength and the divergence from it.

Our simulations have shown that the theoretical maximum detection rate (upper bound) that can be generated at Bob’s site is far beyond the rates that the current technological level can provide. In more detail, due to the hardness of engineering on-demand single-photon sources, most QKD implementations and experiments rely on the use of highly attenuated laser sources which emit probabilistically photons, including also mutli-photon pulses. In order to not compromise security, because of the presence of these multi-photon pulses, the attenuation required is such that the average photon number is set smaller than one, leading thereby to lower detection rates. 
Furthermore, other restrictions of our nowadays technology is the maximum detection rate and the maximum clock rate of the pulses produced by the source; both are much lower than the frequencies indicated in Fig. \ref{fig: Rmax}. However, when technology overcome these problems, the upper bound will be important to be known. Some serious steps have already been done towards this direction, see the works \cite{Dudin,Wang} and future needs in \cite{Reimer}. This upper bound, for maintaining the correct reading order of the symbols sent, could be reached only by the use of single-photon sources in a lossless channel. In case of the probabilistic sources, it cannot be achieved but even in that case this upper bound needs to be fulfilled and the practical use of this theoretical maximum detection rate is presented in Sec. \ref{sec: 7}. 

\begin{figure}[h!]
	\centering
	\includegraphics[height=8.8cm]{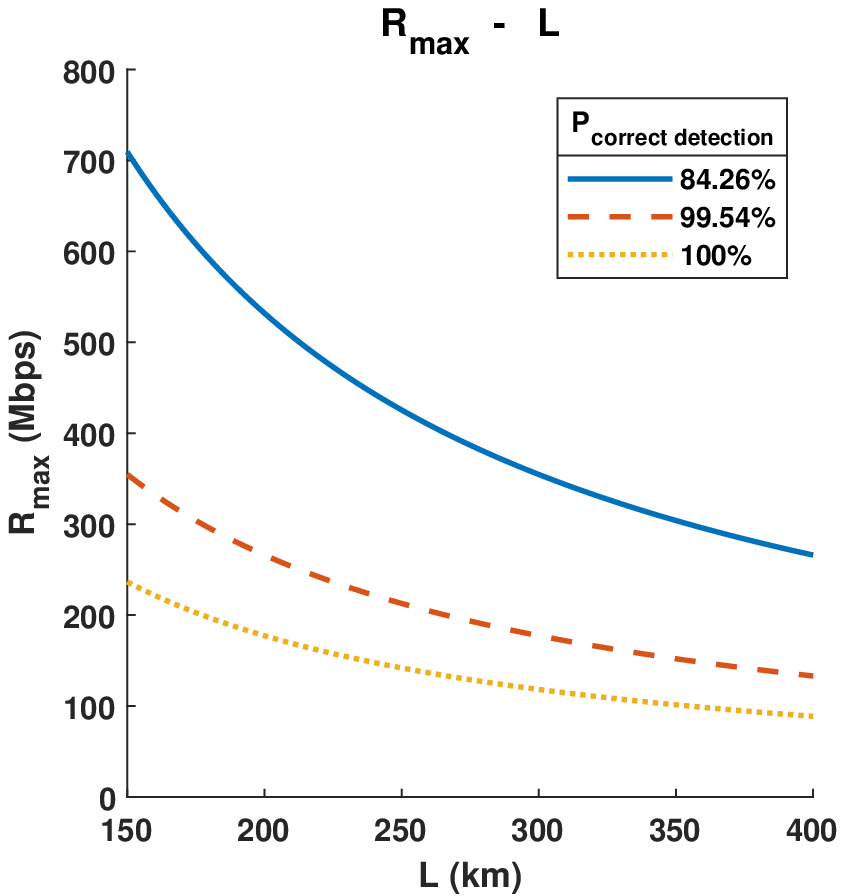}
	\caption{Range of maximum detection rate because of chromatic dispersion for practical devices.}
	\label{fig: Rmax_practical}
\end{figure}

In practical systems, there are single-photon detectors (SPDs) with detection rate in the range of \(100\:MHz\) to \(200\:MHz\) \cite{Practical1} so in Fig. \ref{fig: Rmax_practical} we have chosen to present the maximum achievable distance allowed by chromatic dispersion for the aforementioned detection rate range. In order to achieve longer distances, countermeasures should be taken, some of which we have previously mentioned.

\subsection{The role of the clock rate on the correct ordering reading for dispersive pulse broadening}
\label{sec: 7}
Nowadays, the technology of pulsed laser sources can provide very high clock rates. However, these ultra-high speeds cannot be recorded at the detector's site, due to the fundamental limitations of the semiconductor-based and superconductor-based photon counter technologies (dead and reset time) \cite{deadtime}. Aside from the inherent limitations on the detector's recovery times, the relatively low efficiency and the limited speed of the currently used post-processing error detection algorithms allow for distilled key rates in the order of tens of Mbps in real time deployments \cite{Yuan}. Moving towards longer fiber distances, the photon loss is dramatically increased and is responsible for distilled key rates of few bps even for the Alice pulsed implementations operated on GHz-scale \cite{Geneva}.

However, different steps are needed to be considered depending on the frequency of the clocked setup- because of chromatic dispersion- in order to have even more robust implementations.

From Fig. \ref{fig: Rmax_practical}, we are able to see the upper limit of the theoretical maximum detection rate because of chromatic dispersion. Hence, there are two categories:
\begin{enumerate}
	\item The frequency of the clocked setup is \textit{less} than the theoretical maximum detection rate at a given distance. This condition is easily satisfied for short distances with the current technology or for low frequencies of the clocked setup (recording, inevitably, low detection rate).
	
	\item The frequency of the clocked setup is \textit{higher} than the theoretical maximum detection rate at a given distance. This condition will be greatly boosted in future QKD deployments where high frequencies of the clocked setup will be possible as in \cite{Geneva}.
\end{enumerate}

For each category, we need to follow different strategies for preserving the high-quality reading:

\begin{enumerate}
	\item For the first case, we do not need to compensate the signal using Dispersion Compensating Fiber (DCF); we just need to compute the values of the MZ interferometers' phase-shifters from Eq. \ref{eqn: accuracy} (or see Fig. \ref{fig: Sum of D}). 
	
	Apart from not using DCF, another advantage of this method is its symmetry; permitting two-way communication whilst pre-compensation does not.
	
	\item The second case requires a completely different treatment. If we were using the first strategy for this case, the order of the qubits sent would be changed leading to wrong qubits being read: during the transmittance of the signal, each symbol will be widened to such extent that a part of it will overrun a part of the next symbol leading to, not only intersymbol interference (ISI), but also changed detection order.
	
	Therefore, the use of a dispersion compensation strategy is unavoidable. However, complete compensation of the signal is not necessary. The required steps to identify this minimum compensation length- without the order of the symbols altered- can be found below:
	
	\begin{itemize}
		\item[] We find the maximum active length from Eq. \ref{eqn: R} or Fig. \ref{fig: Rmax} accordingly, which corresponds to the frequency of the clocked setup. Hence, we need to compensate an amount of distance equal to the total length subtracted by this active length. This is the minimum compensation length.
		
		\item[] Of course, we can fully compensate from the beginning of the deployed fiber link at an increased cost since the DCF is much more expensive compared to the standard telecom fibers (e.g. Standard Single Mode Fibers-SSMFs).
	\end{itemize}
	
	For the needs of studying this second case, we considered the implementation settings reported in \cite{Geneva}, where they used a clocked setup at 2.5 GHz. Therefore, from Eq. \ref{eqn: R} or Fig. \ref{fig: Rmax} accordingly, we find the active length equal to 20 km. As a result, instead of performing complete compensation, we need to provide complete compensation for transmission distance equal to \((405-20)\:km=385\:km\). 
	
	Surely, owing to losses the active length can be higher; as most qubits will be lost, the probability for reading with wrong order is non zero but it can be very close to it. Further study is required to emphasize on the actual relation between the active length, the losses and the wrong order reading probability. However, there are protocols such as the User Datagram Protocol (UDP) in which the wrong reading order is not a problem \cite{UDP}.
\end{enumerate}

Although the second case is not symmetrical when pre-compensating, the advantage of not fully compensating the total length can be the cost reduction since the use of longer DCF segments can be prevented.

Overall, we showed two ways- depending on the systems pulse frequency (mentioned as clocked system)- for applying our results and keeping the correct reading order of the transmitted pulses.

These methods can be used not only in this setup with the two MZ interferometers in series but in the general case too; in every fiber optic QKD protocol for which the discrimination of each symbol is important by just changing the Eq. \ref{eqn: R} with:

\begin{equation} \label{eqn: R_general}
\centering
\begin{aligned}
R\leq\dfrac{c}{2X_\varrho}
\end{aligned}		
\end{equation}

The results of this section provide a different interpretation for the results presented on Sec. \ref{sec: 6}; the theoretical maximum detection rate is, also, the maximum clock rate that can be used for maintaining the correct symbol order of the signal sent for a specific transmission distance.

\subsection{Alternative proof for the values of the phase shifter for the phase encoding BB84 QKD protocol} \label{appendix A}
In this section, we demonstrate an alternative way to find the proper choice of the phase shifters using our quantum-mechanical results. The difference between each MZ output lies in the last term of Eq. \ref{position spectra} so this is where we will try to find some indication for the proper choice of the phase shifters.

On its side, the term \(II_{o,p}(x)\) indicates that we can interchange the MZ outputs by changing the sign of its terms; the cosine term is a possible solution for that. Hence, we will try to present the \(z_{ij}-z_{kl}\) term in a more easy to interpret form; from Eq. \ref{more equations final} we have:

\begin{equation} \label{eqn: appendix 1}
\centering
\resizebox{0.49\textwidth}{!}{$\begin{aligned}
z_{ij}(x)-z_{kl}(x)=&\dfrac{1}{\gamma}\left[ k_0(x'_{kl}-x'{ij})+4(\delta k)^4 \delta_1 ({x'_{kl}}^2-{x'_{ij}}^2) \right]\\
=&\dfrac{x'_{kl}-x'{ij}}{\gamma}\left[ k_0+4(\delta k)^4 \delta_1 ({x'_{kl}}+{x'_{ij}}) \right]\\
=&\dfrac{\varDelta_i+\varDelta_j-\varDelta_k-\varDelta_l}{\gamma}\\
& \begin{aligned}
\times \Big[ k_0+ &4(\delta k)^4 \kappa (l_g+2l)\\
&\times (2x-2A_g-A_i-A_j-A_k-A_l)\Big]
\end{aligned}
\end{aligned}$}
\end{equation}
where we have substitute $x'_{ij}$, $x'_{kl}$ and $\delta_1$ to find the third line.

\begin{figure}[h]
	\centering
	\includegraphics[width=\columnwidth]{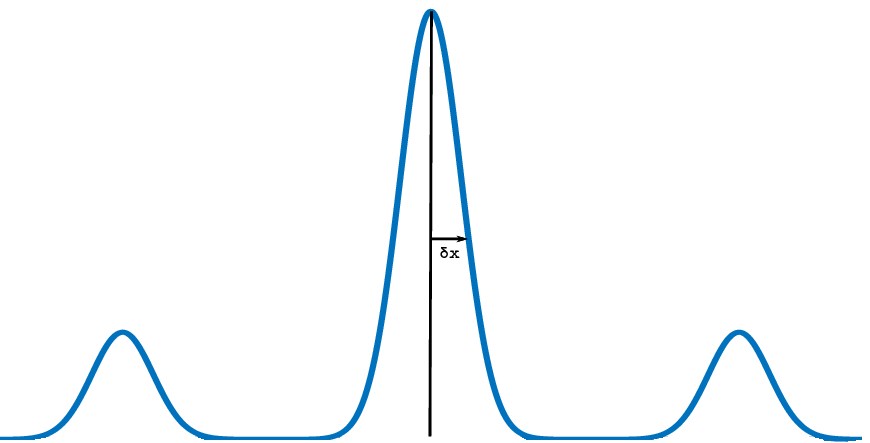}
	\caption{Overview of the axis new zero point. Symbol viewed at the end of the setup (after second Mach-Zehnder).}
	\label{fig: appendix}
\end{figure}

Now, we will do a small trick to reveal the desired form. As we said before the parameter \(\mu_{ij}\) is approximately the same for every pair ij and it represents the distance that the pulse has travelled so far. As a result it would be more convenient to replace the value \(x\) in Eq. \ref{eqn: appendix 1} with the distance that the middle point of the interior pulse will travel. This can be achieved by finding the average value of the minimum and the maximum distance, meaning the short-short travel and the long-long travel of the MZ interferometers. More specifically:

\begin{equation} \label{eqn: appendix 2}
\centering
\begin{split}
x&= \dfrac{\mu_{cc}+\mu_{dm}}{2}+\delta x \\
&= N_0\left( l_g+2l\right)+2\delta_1 k_0 + \dfrac{\varDelta_d+\varDelta_m}{2}+\delta x
\end{split}
\end{equation}
where \(\delta x\) has been added to maintain the meaning of \(x\). See Fig. \ref{fig: appendix} for the visualisation of the meaning of \(\delta x\).

Substituting Eq. \ref{eqn: appendix 2} to Eq. \ref{eqn: appendix 1}, we get:

\begin{equation} \label{eqn: appendix 3}
\centering
\begin{aligned}
&z_{ij}(\delta x)-z_{kl}(\delta x)=\dfrac{\varDelta_i+\varDelta_j-\varDelta_k-\varDelta_l}{\gamma} \\
& \begin{aligned}
\times \bigg\{  k_0+&4(\delta k)^4 \kappa (l_g+2l)\\
& \begin{aligned}
\times \Big[2\delta x&+ 4\delta_1 k_0 +\varDelta_d +\varDelta_m \\
&-(\varDelta_i+\varDelta_j+\varDelta_k+\varDelta_l) \Big] \bigg\}
\end{aligned}
\end{aligned}
\end{aligned}
\end{equation}

Substituting \(\gamma\) and after a few steps we get:

\begin{equation} \label{eqn: appendix 4}
\centering
\begin{split}
&z_{ij}(\delta x)-z_{kl}(\delta x)=\dfrac{2\pi}{\lambda_0}\left( \varDelta_i+\varDelta_j-\varDelta_k-\varDelta_l\right)\\
& \begin{aligned}
\times &\bigg\{ 1+\dfrac{\lambda_0 (1-\frac{1}{\gamma})}{4\pi \kappa (l_g+2l)} \\
&\times \Big[ \delta x +\dfrac{\varDelta_d +\varDelta_m- (\varDelta_i+\varDelta_j+\varDelta_k+\varDelta_l)}{2}\Big] \bigg\}
\end{aligned}
\end{split}
\end{equation}

Hereinafter, we will call the term $\frac{\lambda_0 (1-\frac{1}{\gamma})}{4\pi \kappa (l_g+2l)}$, $G$. When there is no chromatic dispersion $G=0$ and also in case that the second term of Eq. \ref{eqn: appendix 4} is much smaller than 1, it can be easily proved that chromatic dispersion has no significant effect in the bases choice.

\begin{figure}[h!]
	\centering
	\includegraphics[height=8.8cm]{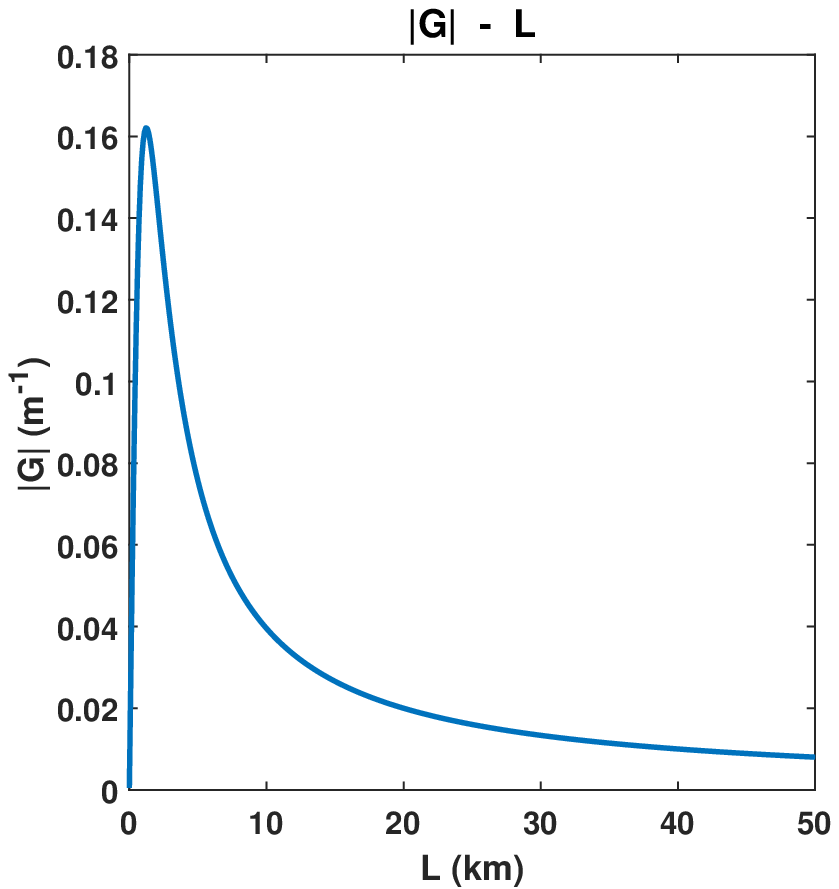}
	\caption{The absolute value of the term G in relevance with the transmission distance (L). G reaches its maximum value around 1.236 km. Simulations showed that different lengths of the MZ legs have no impact on the value of G (here $l=1\:m$) and that for greater values of transmission distances G tends to zero.}
	\label{fig:G_L}
\end{figure}

It is very interesting that the term $G$ presents a maximum value when the transmission distance is around $1.236\:km$ as shown in Fig. \ref{fig:G_L}. This maximum value is not insignificant and thus the whole second term of \ref{eqn: appendix 4} needs to be investigated.

The middle pulse that we are interested in is represented by the term \(z_{cm},z_{dc}\), so the term \(\varDelta_i+\varDelta_j-\varDelta_k-\varDelta_l\) becomes \(\varDelta_m-\varDelta_d\); the sign is not important due to the cosine that follows. Equation \ref{eqn: appendix 4} indicates that only the difference between the values of \(\varDelta\) is expected to affect the final result. Hence, we will set:

\begin{equation} \label{eqn: appendix 6}
\centering
\begin{aligned}
\varDelta_{\omega}=\widetilde{\varDelta}+\phi_{\omega} & & \text{where \(\omega\in\{m,d\}\)}
\end{aligned}
\end{equation}
where we used the letter \(\phi\), which is normally used for phase values and is of the order of wavelength, to keep in mind that we are looking for the phase correspondence.

Hence Eq. \ref{eqn: appendix 4} becomes:

\begin{equation} \label{eqn: appendix new1}
\centering
\begin{split}
z_{cm}(\delta x)-z_{dc}(\delta x)=\dfrac{2\pi}{\lambda_0}\left( \phi_m-\phi_d\right) \bigg\{ 1+G\Big[ \delta x - \varDelta_c\Big] \bigg\}
\end{split}
\end{equation}

Three main results can be extracted from Eq. \ref{eqn: appendix new1}:
\begin{enumerate}
	\item $\varDelta_c$ displace the point where the outputs are interchanged as expected from Eq. \ref{eqn: difference}. Although usually $\varDelta_c=0$.
	
	\item When $\delta x$ is $0$, which corresponds to the middle of the middle pulse, the effect of chromatic dispersion is negated. However, it is not known yet, the effect of chromatic dispersion in every other point of the middle pulse. Thus, the absolute value of $G\cdot \delta x$ needs to be investigated and the value $\delta x=3\cdot\sigma$ is the best to be chosen as it corresponds to the end of the middle pulse.
\end{enumerate}

\begin{figure}[h!]
	\centering
	\includegraphics[height=8.8cm]{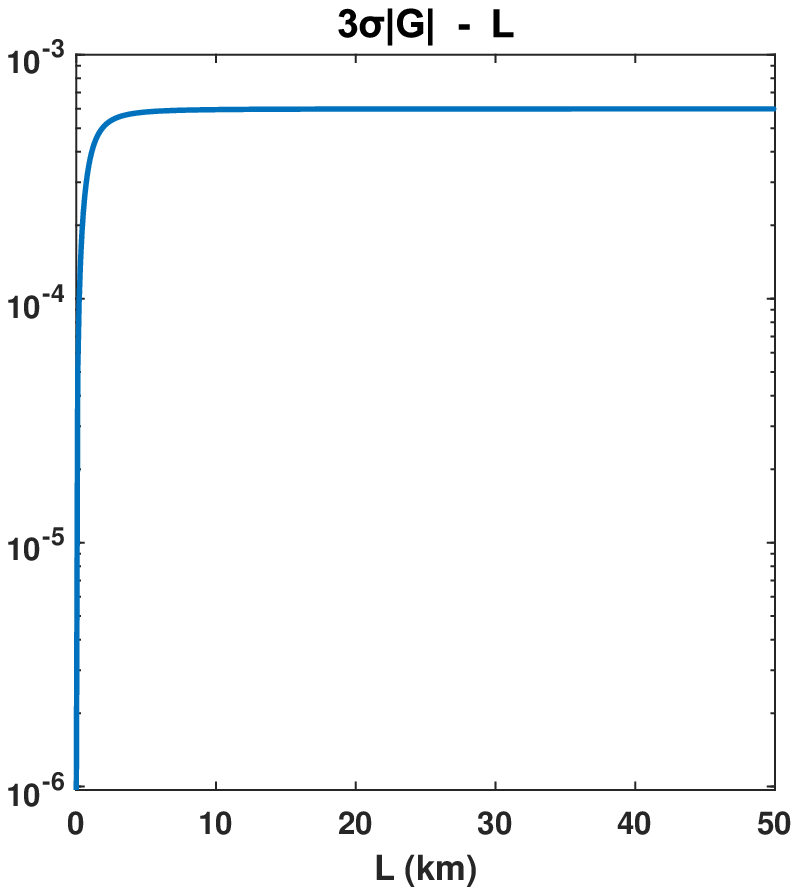}
	\caption{The absolute value of the second term of Eq. \ref{eqn: appendix new1} in relevance with the transmission distance (L). The second term reaches its maximum value around 2 km. Simulations showed that different lengths of the MZ legs have no impact on its value (here $l=1\:m$) and that for greater values of transmission distances it tends to a plateau. The y-axis is logarithmic and it has no units (pure number).}
	\label{fig:G_dx}
\end{figure}

In Fig. \ref{fig:G_dx} the absolute value of the aforementioned second term for different transmission distances is indicated.

Therefore, it is safe to make the following assumption:

\begin{equation} \label{eqn: appendix 5}
\centering
\begin{aligned}
(z_{cm}(\delta x)-z_{dc}(\delta x))|_{\delta x=0} &\simeq\dfrac{2\pi}{\lambda_0}\left( \phi_m-\phi_d\right)
\end{aligned}
\end{equation}

Now, we have to assume what phases we should pick for the creation of the bases. Phase encoding make use of two pairs of phases that each signal pulse should be modulated in. There is need to:

\begin{enumerate}
	\item For the same pair, the phases need to be chosen so that they will lead to orthogonal states creation; maximize the detection probability to different detectors.
	
	\item When read in wrong basis, the result is ambiguous.
\end{enumerate}

We will start searching by setting for convenience these two values equal to zero; meaning \(\phi_m=\phi_d=0\), so the cosine term is precisely equal to 1. Now, for this choice we need to change only the \(\phi_d\) so as to invert the outputs. It is easy to find that we can accomplish this by choosing \(\phi_d=\frac{\lambda_0}{2}\), which corresponds to a phase equal to \(\pi\), so, now, the cosine is approximately equal to -1 (instead of 1), indicating that the outputs were inverted.

So far, we have defined the first basis. We also need to find the second one. It must be chosen such that if Bob reads in respect to the second one but Alice has modulated the qubit using the first one, Bob will not be able to decode any information.

A reasonable choice for this to happen is by choosing the second basis to be modulated by adding a phase equal to \(\frac{\pi}{2}\); meaning \(\phi_m=\frac{\lambda_0}{4}\). We can see that in this case if Alice sends a symbol which has been modulated using the first basis (\(\phi_d=0\:or\:\phi_d=\frac{\lambda_0}{2}\)) and Bob reads this symbol by applying \(\phi_m=\frac{\lambda_0}{4}\), the cosine for both cases are approximately zero, thereby they have the same detection probability to each output.

By the same logic, we can see that Alice needs to modulate applying \(\phi_d=\frac{\lambda_0}{4}\:or\:\phi_d=\frac{3\lambda_0}{4}\) in order for Bob to be able to read when \(\phi_m=\frac{\lambda_0}{4}\).

At last, we have find the two bases that we wanted and our results can be confirmed by the simulations in the next section. Furthermore, the insignificant effect of chromatic dispersion on the choice of the bases was proved. Summing up, we have that Alice creates the qubits by applying:
\begin{equation}
\phi_d= \left\{
\begin{array}{llllllll} \label{table: appendix 1}
0 && or && \frac{\lambda_0}{2} & & &\text{, basis X}\\
\\
\frac{\lambda_0}{4} && or && \frac{3\lambda_0}{4} & & & \text{, basis Z}\\
\end{array} 
\right. 
\end{equation}

where the first value from each row, usually corresponds to bit 0 and the second one to bit 1. Bob reads the qubits by applying:
\begin{equation}
\phi_m= \left\{
\begin{array}{llll} \label{table: appendix 2}
0 & & &\text{, basis X}\\
\\
\frac{\lambda_0}{4} & & & \text{, basis Z}\\
\end{array} 
\right. 
\end{equation}

\section{Example: BB84 QKD protocol simulation based on our results}
\label{sec: 6}
Before proceeding, we need to highlight the main results of this study, which are presented in Eqs. \ref{eqn: accuracy} or \ref{eqn: accuracy2}, \ref{eqn: R} or \ref{eqn: R2} and in \ref{eqn: synchronization}.

To make these results more concrete, we will apply them on the phase encoding BB84 QKD protocol. Obviously, the same procedure applies also on the phase encoding decoy-state QKD protocol, as it uses the same logic for the construction of the qubits, thereby the results presented in this subsection are true for both protocols.

Here, we assume the same characteristics of the source as mentioned in Sect. \ref{sec: 5} and we want to accomplish a communication over the \(50\:km\) distance with as less error detection as possible, meaning \(\varrho=3\) from table \ref{table:k and probability}. We use Eqs. \ref{eqn: accuracy} or \ref{eqn: accuracy2}. We will, also, assume that \(t_{rising}=t_{falling}=0\: s\), so we obtain:

\begin{equation} \label{eqn: ex}
\varDelta_d+\varDelta_m \geq 0.423\: m
\end{equation}

In practical designs, the value of \(0.5\:m\) would be chosen for safety reasons, as it is higher than the limit value, leading to a relatively more robust system to the effects emerging from plausible other imperfections, and also because any divergence from this value will keep us within bounds. This means that whatever choice we have to make in order to create the necessary pair of bases for these protocols, the condition presented in Eq. \ref{eqn: ex} must be always true. Hence, it is reasonable to assume a minimum value equal to \(0.5/2=0.25\:m\) for each of the two phase shifters.

In Sec. \ref{appendix A} we extracted a simple way to understand the choice of the value of the two phase shifters:

\begin{enumerate}
	\item Bob chooses to read randomly between the two bases, by setting \(\phi_m=0\) or \(\phi_m=\frac{\lambda_0}{4}\) for basis X and basis Z respectively.
	
	\item Alice modulates in basis X when the values of phase shifters are \(\phi_d=0\) or \(\phi_d=\frac{\lambda_0}{2}\) for bits 0 and 1 respectively. Whilst, she modulates in basis Z when the values of phase shifters are \(\phi_d=\frac{\lambda_0}{4}\) or \(\phi_d=\frac{3\lambda_0}{4}\) for bits 0 and 1 respectively.
\end{enumerate}
For the meaning of the symbol \(\phi_{\omega}\) where \(\omega\in\{d,m\}\) see Eq. \ref{eqn: appendix 6} in Sec. \ref{appendix A}.

Of course, it should be clear by now, that the aforementioned values are added to the value 0.25 m that we extracted before by adding in series a constant length (0.25 m) to the phase modulator: \(\varDelta=(0.25\:m)+\phi_{\omega}\) where \(\phi_{\omega}\) adds the phase modulation which corresponds to $0,\:\frac{\lambda_0}{4},\:\frac{\lambda_0}{2}\\ \:or\:\frac{3\lambda_0}{4}$.

\begin{figure}[h!]
	\begin{subfigure}[t]{0.48\columnwidth}\centering
		\includegraphics[width=\linewidth]{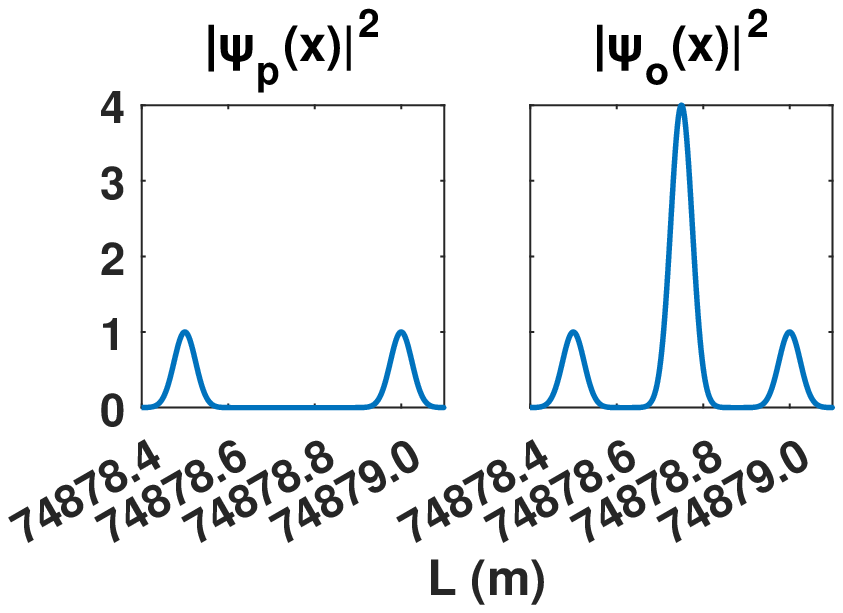}
		\caption{\textbf{\boldmath\(\phi_d=0\)}}
		\label{fig: dd0}
	\end{subfigure}
	\begin{subfigure}[t]{0.48\columnwidth}\centering
		\includegraphics[width=\linewidth]{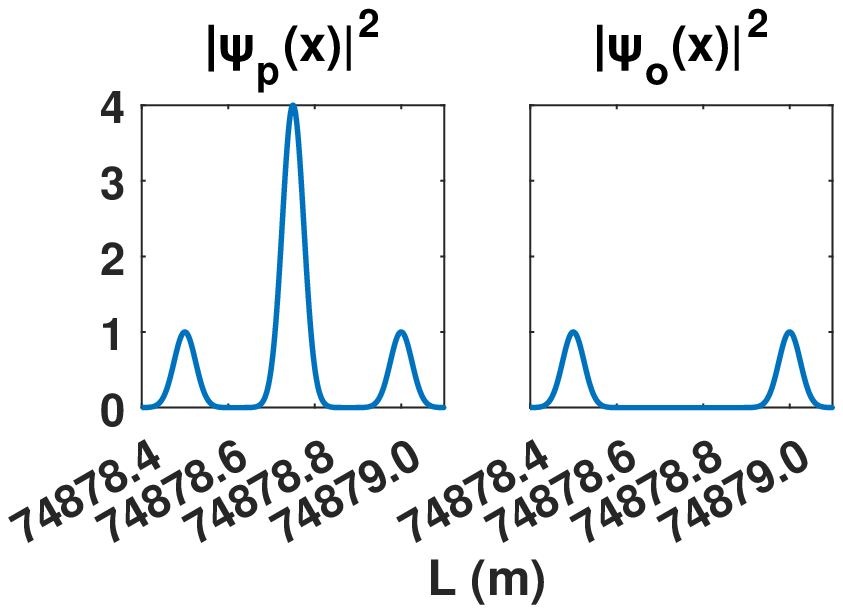}
		\caption{\boldmath\(\phi_d=\frac{\lambda_0}{2}\)}
		\label{fig: dd2}
	\end{subfigure}
	\par\bigskip
	\begin{subfigure}[t]{0.48\columnwidth}\centering
		\includegraphics[width=\linewidth]{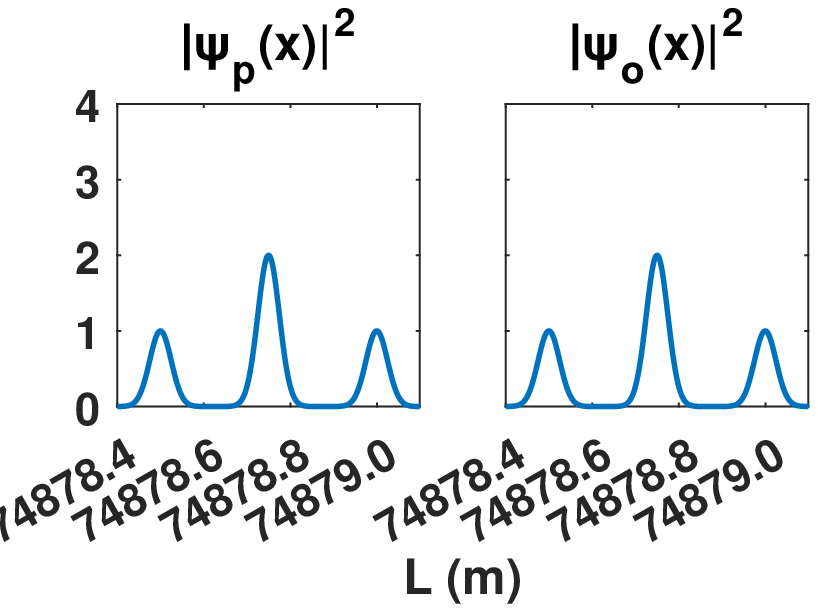}
		\caption{\boldmath\(\phi_d=\frac{\lambda_0}{4}\)}
		\label{fig: dd4}
	\end{subfigure}
	\begin{subfigure}[t]{0.48\columnwidth}\centering
		\includegraphics[width=\linewidth]{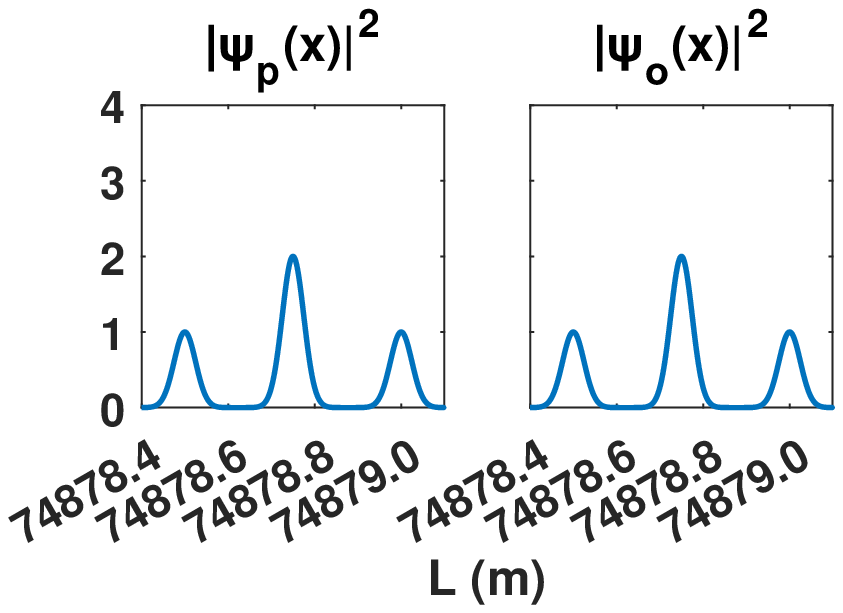}
		\caption{\boldmath\(\phi_d=\frac{3\lambda_0}{4}\)}
		\label{fig: 3dd4}
	\end{subfigure}
	\caption{Position Spectra at the MZ outputs (o,p) for \(\phi_m=0\) and every value of \(\phi_d\). Bob reads on basis X. The \(\phi_m\), \(\phi_d\) represent the phase shift values of Bob and Alice respectively. The y-axis is normalized for the relative height difference of the pulses to be directly visible.}
	\label{fig: delta}
\end{figure}

\begin{figure}[h!]
	\begin{subfigure}[t]{0.48\columnwidth}\centering
		\includegraphics[width=\linewidth]{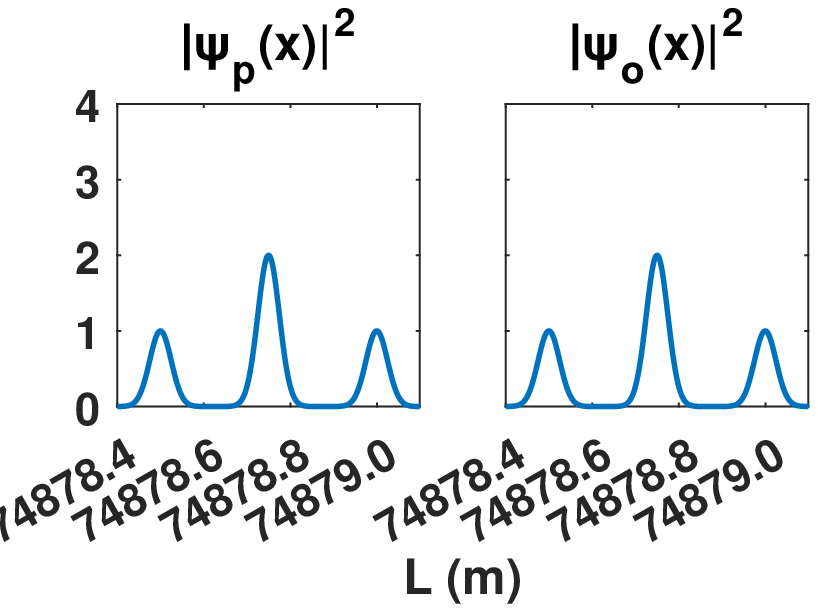}
		\caption{\boldmath\(\phi_d=0\)}
		\label{fig: dd0_2}
	\end{subfigure}
	\begin{subfigure}[t]{0.48\columnwidth}\centering
		\includegraphics[width=\linewidth]{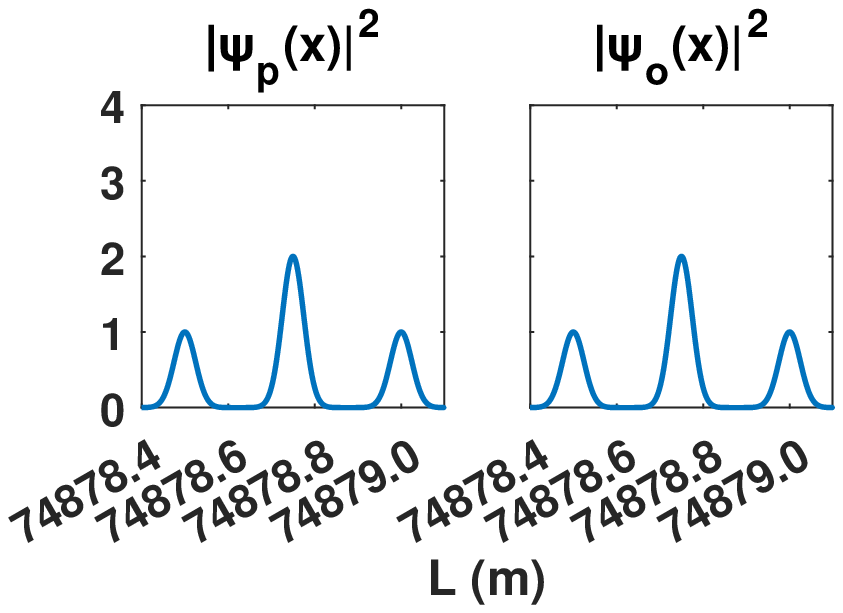}
		\caption{\boldmath\(\phi_d=\frac{\lambda_0}{2}\)}
		\label{fig: dd2_2}
	\end{subfigure}
	\par\bigskip
	\begin{subfigure}[t]{0.48\columnwidth}\centering
		\includegraphics[width=\linewidth]{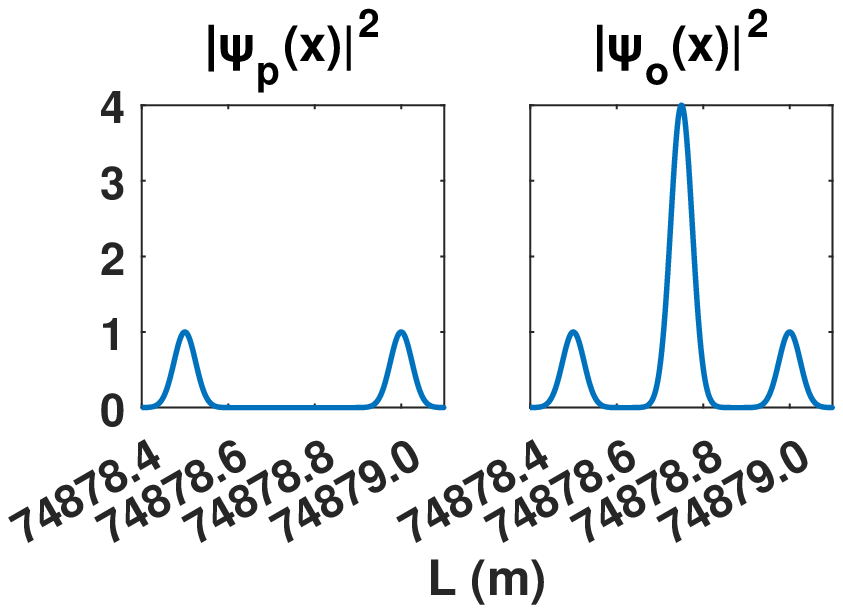}
		\caption{\boldmath\(\phi_d=\frac{\lambda_0}{4}\)}
		\label{fig: dd4_2}
	\end{subfigure}
	\begin{subfigure}[t]{0.48\columnwidth}\centering
		\includegraphics[width=\linewidth]{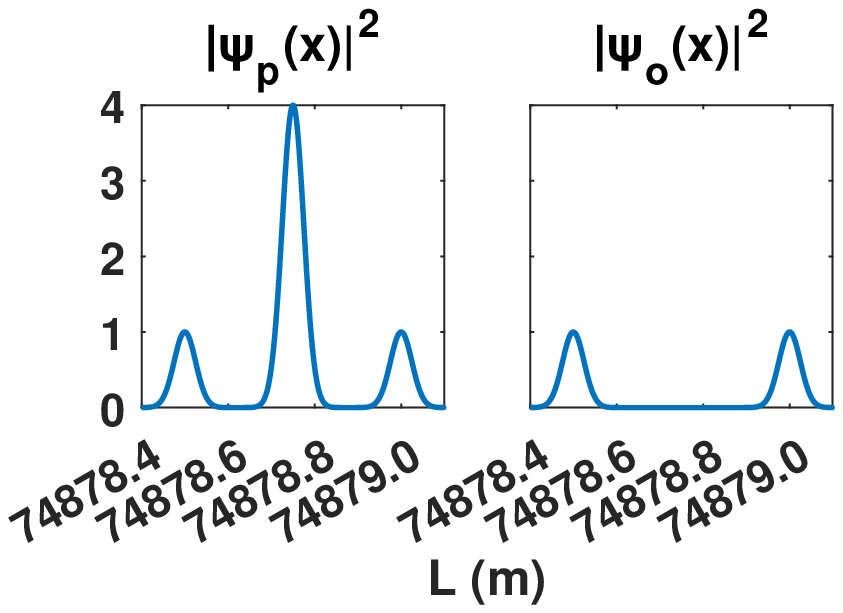}
		\caption{\boldmath\(\phi_d=\frac{3\lambda_0}{4}\)}
		\label{fig: 3dd4_2}
	\end{subfigure}
	\caption{Position Spectra at the MZ outputs (o,p) for \(\phi_m=\frac{\lambda_0}{4}\) and every value of \(\phi_d\). Bob reads on basis Z. The \(\phi_m\), \(\phi_d\) represent the phase shift values of Bob and Alice respectively. The y-axis is normalized for the relative height difference of the pulses to be directly visible.}
	\label{fig: delta_2}
\end{figure}

Our results can be confirmed by precise simulations, see Figs. \ref{fig: delta} and \ref{fig: delta_2}. In these figures, we are able to see the importance of the distance between two consecutive symbols and, thus the importance of the results in Sec. \ref{sec: 5}. If we had not considered the minimum values of the phase shifters indicated by Eq. \ref{eqn: accuracy}, the middle pulse would not be zero in the cases that it has to be (Figs. \ref{fig: dd0}, \ref{fig: dd2}, \ref{fig: dd4_2} and \ref{fig: 3dd4_2}).
\FloatBarrier
\bigskip

Furthermore, from Eq. \ref{eqn: R} or \ref{eqn: R2} we obtain that the maximum possible detection rate that can be generated without countermeasures is approximately \(710\:Mbps\) and \(473\:Mbps\) respectively.
Finally, we are able to find from Eq. \ref{eqn: synchronization} the maximum duration of the detector's window: \(\Delta t_{gate}=0.962\:ns\).

In a similar way, our results can be used for any distance with non-ideal detectors and for any QKD protocol that uses the same two Mach-Zehnder setup; they can be easily expanded to other setups as mentioned in Eq. \ref{eqn: R_general}.


\section{Conclusions}
\label{sec: 8}
Maximum readability QKD systems are necessary to reach detection rates as high as possible, where \(100\%\) visibility -without any other imperfection other than chromatic dispersion- is needed to achieve that. We have found that in order to achieve this probability the detection window of the detector's gate should be \(7.206\cdot FWHM\); in case that the detector's window has non-ideal rising and falling times, a safety factor should be added to the aforementioned value.

Apart from the width of the detector's window, as a result of chromatic dispersion, the MZ phase shift values should also be appropriately chosen in order to attain high visibility. Lower bound for the sum of the fiber length of phase shifters of the two MZ interferometers has been calculated and, in high precision, it was found to exhibit linear dependence on the fiber transmission length. More specifically, the slope of the dependence is approximately equal to \(0.8454\: \frac{m}{100\:km}\) for achieving \(100\%\) visibility.

The lower bound restriction of the MZ interferometers phase shifters leads to an upper bound restriction for the maximum detection rate that can be recorded because of chromatic dispersion in the setup. The upper bound depends on the ISI appeared between two consecutive pulses and it follows an inversely proportional relation with the transmission length. More specifically, the constant of proportionality of this inverse dependence is, approximately, equal to \(35.46\:Gbps\cdot km\) for long transmission distances for which we are interested.

In contrast to the ISI effects in classical flows, there is the need to consider the intended visibility associated with the setup too and the aforementioned value (\(35.46\:Gbps\cdot km\)) is for accomplishing \(100\%\) visibility. Besides the role of ISI effect, the upper bound can be also affected by the quantum non-linear photon-to-photon interaction and this mechanism should also be considered as part of our study. In this case a constant of proportionality of the aforementioned inverse dependence equal to \(11.82\:Gbps\cdot km\) is derived. In this line, our results have taken this into consideration and they are valid either way.

Finally, depending on the pulse generation frequency, the upper bound can be used to select the most efficient compensation scheme to ensure the correct order of the signal and greatly reducing the deployment cost of the fiber-based setup. These methods can be easily expanded to QKD protocol implementations where the discrimination of each symbol is an essential parameter- to the best of our knowledge, every known protocol makes use of that; this expansion can happen by changing the constant of proportionality of the inverse dependence from \(35.46\:Gbps\cdot km\) to \(70.92\:Gbps\cdot km\).

Welcoming the beyond-500 km era for secure distance of QKD in fiber links \cite{new_era}, our work contribute in this topic of ultra-long haul, repeaterless QKD transmission where the role of readability, time synchronization and chromatic dispersion of the QKD implementation should be carefully addressed.

\section*{Acknowledgement}
We thank the two anonymous reviewers for their comments and suggestions which strongly improved and clarify our work.

We, also, thank Constantine Nisidis for the creation of the graphical abstract.

\section*{Authors' contributions}
\label{sec: Authors contributions}
C. Papapanos is the corresponding author of this research. C. Papapanos conceived and comprehended the presented idea and developed the theory as well as the analytical calculations. D. Zavitsanos, A. Raptakis and G. Giannoulis were, also, involved supporting the literature survey and the preparation of the manuscript. All the authors have read and approved the final manuscript.

\begin{appendices}
\numberwithin{equation}{section}
\section{Pre-compensation, symmetric compensation and post-compensation effect when DCF is used} \label{appendix B}

First, the pre-compensation case is studied where the DCF is added in Alice's side. This will change the Gaussian pulse sent by Alice to:

\begin{equation} \label{eqn:appendix_2_1}
	\alpha'_{a}(k)=\alpha_a(k)\sqrt{T_{cp}} exp[-ikA_{cp}-ik^2B_{cp}]
\end{equation}
where $\alpha_a(k)$ is the Gaussian pulse before the pre-compen-sation and is given by \cite{CD}:
\begin{equation} \label{eqn:appendix_2_2}
\alpha_{a}(k)=[2\pi (\delta k)^2]^{-1/4}exp\{-\dfrac{(k-k_0)^2}{4(\delta k)^2}\}
\end{equation}

All the other symbols in Eqs. \ref{eqn:appendix_2_1} and \ref{eqn:appendix_2_2} have the same meaning as in the main text. The index $cp$ signifies that the parameters are produced by the compensation; thus $A_{cp}=N_0l_{cp}$, $B_{cp}=\kappa_{cp}l_{cp}$ and $T_{cp}$ are the losses emerged from a DCF of length equals to $l_{cp}$.

Hereinafter, we derive the output of the setup using the methodology of \cite{CD}. Hence, after some calculation the wave number functions of the outputs can be expressed:

\begin{equation}  \label{eqn:appendix_2_3}
	\resizebox{0.5\textwidth}{!}{$\begin{aligned}
		&\alpha_{o,p}(k)=\frac{1}{4}\Bigg[\sqrt{T_gT_{cp}} exp\{ -ik(A_g+A_{cp})-ik^2(B_g+B_{cp}) \} \\     
		&\times \big(\sqrt{T_m}exp\{ -ikA_m-ik^2B_m \}\pm\sqrt{T_n}exp\{ -ikA_n-ik^2B_n \}\big) \\
		&\times \big(\sqrt{T_d}exp\{ -ikA_d-ik^2B_d \}-\sqrt{T_c}exp\{ -ikA_c-ik^2B_c \}\big)\\
		&\times \alpha_{a}(k)\Bigg]
		\end{aligned}$}
\end{equation}

Hence, the output is similar to the output derived in \cite{CD} with the differences being:
\begin{enumerate}
	\item $A_g$ is replaced by $A_g'=A_g+A_{cp}=N_0l_{cp}+\varDelta_g+N_0l_g$.
	\item $B_g$ is replaced by $B_g'=B_g+B_{cp}=\kappa l_g+\kappa_{cp}l_{cp}$ .
	\item There is an added loss equal to $T_{cp}$ which can be unified with the $T_g$ by replacing the latter with $T_g'=T_gT_{cp}$ .
\end{enumerate} 

Compensating demands choosing appropriate values for the parameters of the DCF such that $\delta_1=0$ (defined in Eq. \ref{delta}), i.e. $\kappa_{cp}l_{cp}=-\kappa (l_g+2l)$, the compensated output position spectra is found:

\begin{equation} \label{eqn:appendix_2_4}
\centering
\begin{split}
|\psi_{o,p}(x)|^2 =&\dfrac{T_g'}{32\pi\sqrt{2\pi}(\delta k)} \{|J_{dm}(x)|^2+|J_{cm}(x)|^2\\
&+|J_{dc}(x)|^2+|J_{cc}(x)|^2+2II_{o,p}(x) \}
\end{split}
\end{equation}

where $II_{o,p}(x)$ is the same as in Eq. \ref{II} but the single quantities are given by:

\begin{equation} \label{eqn:appendix_2_5}
\centering
\begin{split}
|J_{ij}(x)|^2 =& 4\pi (\delta k)^2 T^2 exp\{-2(\delta k)^2[x-\mu_{ij}]^2\}\:,
\\		
\Re[J_{ij}(x) J^*_{kl}(x)] =& 4\pi(\delta k)^2T^2 exp\{-2(\delta k)^2[x-\mu_{ij}]^2\}\\
&\times cos[k_0(\varDelta_k+\varDelta_l-\varDelta_i-\varDelta_j)]
\end{split}
\end{equation}

where $\mu_ij$ is given from Eq. \ref{eqn: mean value} with $\delta_1=0$ and also we have assumed, as in the main text, that $l_c=l_d=l_m=l_n$.

Comparing the position spectra at the beginning of the setup and before the pre-compensation given by:
\begin{equation} \label{eqn:appendix_2_6}
\centering
	|\psi_a(x)|^2=\sqrt{\dfrac{2(\delta k)^2}{\pi}} exp\{-2(\delta k)^2 x^2\}
\end{equation}
with Eq. \ref{eqn:appendix_2_4} is easily revealed that the effect of chromatic dispersion has been compensated.

The same methodology will end to similar final equations for the cases of post-compensation and symmetric compensation, when DCF is used, concluding to the same result; the aforementioned compensation techniques do not alter the main text's results.
\end{appendices}
\end{document}